\documentclass[preprint,12pt]{elsarticle}

\usepackage{graphicx}

\usepackage[table,xcdraw]{xcolor}

\usepackage{amssymb}
\usepackage{amsmath}
\usepackage{url}
\usepackage{lineno}

\biboptions{authoryear}

\makeatletter
\def\ps@pprintTitle{%
 \let\@oddhead\@empty
 \let\@evenhead\@empty
 \def\@oddfoot{}%
 \let\@evenfoot\@oddfoot}

\makeatother

\begin{document}

\begin{frontmatter}

\title{Multifractal cross-correlations between the World Oil and other Financial Markets in 2012-2017}

\author[ifj]{Marcin W\c atorek\corref{cor1}}
\ead{marcin.watorek@ifj.edu.pl}
\cortext[cor1]{Corresponding author}
\author[ifj,pk]{Stanis{\l}aw Dro\.zd\.z}
\author[ifj]{Pawe{\l} O\'swi\c ecimka}
\author[pk]{Marek Stanuszek}

\address[ifj]{Complex Systems Theory Department, Institute of Nuclear Physics, Polish Academy of Sciences, ul.~Radzikowskiego 152, 31-342 Krak\'ow, Poland}
\address[pk]{Faculty of Physics, Mathematics and Computer Science, Cracow University of Technology, ul.~Warszawska 24, 31-155 Krak\'ow, Poland}

\begin{abstract}

Statistical and multiscaling characteristics of WTI Crude Oil futures prices expressed in US dollar in relation to the most traded currencies as well as to gold futures and to the E-mini S\&P500 futures prices on 5 min intra-day recordings in the period January 2012 - December 2017 are studied. It is shown that in most of the cases the tails of return distributions of the considered financial instruments follow the inverse cubic power law. The only exception is the Russian ruble for which the distribution tail is heavier and scales with the exponent close to 2. From the perspective of multiscaling the analysed time series reveal the multifractal organization with the left-sided asymmetry of the corresponding singularity spectra. Even more, all the considered financial instruments appear to be multifractally cross-correlated with oil, especially on the level of medium-size fluctuations, as the multifractal cross-correlation analysis carried out by means of the multifractal cross-correlation analysis (MFCCA) and detrended cross-correlation coefficient $\rho_q$ show. The degree of such cross-correlations is however varying among the financial instruments. The strongest ties to the oil characterize currencies of the oil extracting countries. Strength of this multifractal coupling appears to depend also on the oil market trend. In the analysed time period the level of cross-correlations systematically increases during the bear phase on the oil market and it saturates after the trend reversal in 1st half of 2016. The same methodology is also applied to identify possible causal relations between considered observables. Searching for some related asymmetry in the information flow mediating cross-correlations indicates that it was the oil price that led the Russian ruble over the time period here considered rather than vice versa.  

\end{abstract}

\end{frontmatter}

\section{Introduction}
\label{Intro}

Oil constitutes one of the most demanded commodities in the world and it is largely for this reason that the crude oil market is bigger than all raw metal markets combined. Its size at current prices is \$1.7 trillion per year \citep{desjardins2016}. In earliest economic studies \cite{Golub83} and \cite{Krugman83}, pointed out that an oil-exporting (oil-importing) countries may experience exchange rate appreciation (depreciation) when oil prices rise (fall). It also is clear that oil prices affect all aspects of world economy and politics \citep{Ratti2015,Ratti2016}.  

Because oil is quoted in dollars, there are natural relations between oil and currencies pointed by \cite{Sadorsky2000}, \cite{Akram2004}, \cite{Benassy2007}, \cite{Reboredo2012}, \cite{Turhan2014} and \cite{Beckman2016}. Various factors shaping correlations between the oil and stock markets has been studied in \cite{Jones1996}, \cite{Sadorsky99}, \cite{Basher2012}, \cite{Sadorsky2012}, \cite{Zhang2016}, \cite{Balcilar2017} and \cite{Reboredo2017}. The oil market thus constitutes an inseparable component of the global world financial system and, as such, it crucially influences its complexity characteristics \citep{kwapien2012}. Description of the related dependences and correlations seems therefore to be the most natural by using methodology of time series analysis that allows to take care of the nonlinear effects and that has already proved fruitful in many domains of complexity, including the financial markets \citep{mandelbrot1997,calvet2002,Bouchaud2004,Grech2004,lux2008,
Morales2012,Rak2015,Gubiec2017,Jiang2018,Klamut2018}. This methodology includes the Wavelet Transform Modulus Maxima - WTMM \citep{Muzy91}, and a series of methods based on detrending with an increasing degree of generality. These are the Detrended Fluctuation Analysis - DFA \citep{kantelhardt01}, the Detrended Cross-Correlation Analysis - DCCA \citep{podobnik2008} and their multifractal generalizations: Multifractal Detrended Fluctuation Analysis - MFDFA \citep{kantelhardt02}, Multifractal Detrended Cross-Correlation Analysis - MFDCCA \citep{zhou2008}, Multifractal Detrending Moving-Average Cross-Correlation Analysis - MFXDMA \citep{Jiang2011} and Multifractal Cross-Correlation Analysis - MFCCA \citep{oswiecimka2014}. Multifractal cross-correlations can also be studied within the wavelet formalism by making use of the Multifractal Cross Wavelet Transform (MF-X-WT) analysis \citep{Jiang2017}.

There already are several contributions presenting application of the detrended methods to investigate correlations between the oil and currency markets \citep{Pal2014,Reboredo2014,Li2016,Hussain2017} or to the stock market \citep{Wang2012,Ma2013,Ma2014,Yang2016,Ferreira2019}. Thus far only daily data were used for those analyses. 

The speed of information processing in contemporary markets is systematically increasing and therefore in order to better capture the underlying dynamics the present analysis is performed for higher frequency intraday 5 min recordings in the period between January 02, 2012 and December 29 2017. This period is particularly interesting because of the positive bubble on the US dollar and the negative bubble on the oil market as reported by \cite{Tokic2015}, \cite{Fantazzini2016}, \cite{Fomin2016} and \cite{Watorek2016}. In the present study the MFCCA proposed by \cite{oswiecimka2014} and the $q$-dependent detrended cross-correlation coefficient $\rho_q$ proposed by \cite{kwapien2015} are employed to analyse the cross-correlations between WTI Crude Oil futures (CL) and thirteen most important financial instruments:  E-mini S\&P500 futures (ES) as an appropriate representation of the world stock market, gold futures (GC) and 11 currencies expressed in the US dollar. The currencies thus include Australian dollar (AUD/USD), Canadian dollar (CAD/USD), Renminbi (CNH/USD), Euro (EUR/USD), Pound sterling (GBP/USD), Japanese yen (JPY/USD), Mexican peso (MXN/USD), Norwegian krone (NOK/USD), Polish zloty (PLN/USD),  Russian ruble (RUB/USD) and South African rand (ZAR/USD). Time development of all those financial instruments over the time period considered is shown in Fig. \ref{fig:CL}. 

In the present paper the following questions are addressed:
(i) What is the cross-correlation level between oil and other financial instruments?
(ii) Are these cross-correlations multifractal?
(iii) Are the identified relations stable in sub-periods?
(iv) Is there asymmetry in cross-information flow between oil and other financial instruments?

\section{Data specification}
\label{data}

\begin{figure}[h!]
\centering 
\includegraphics[scale=0.45]{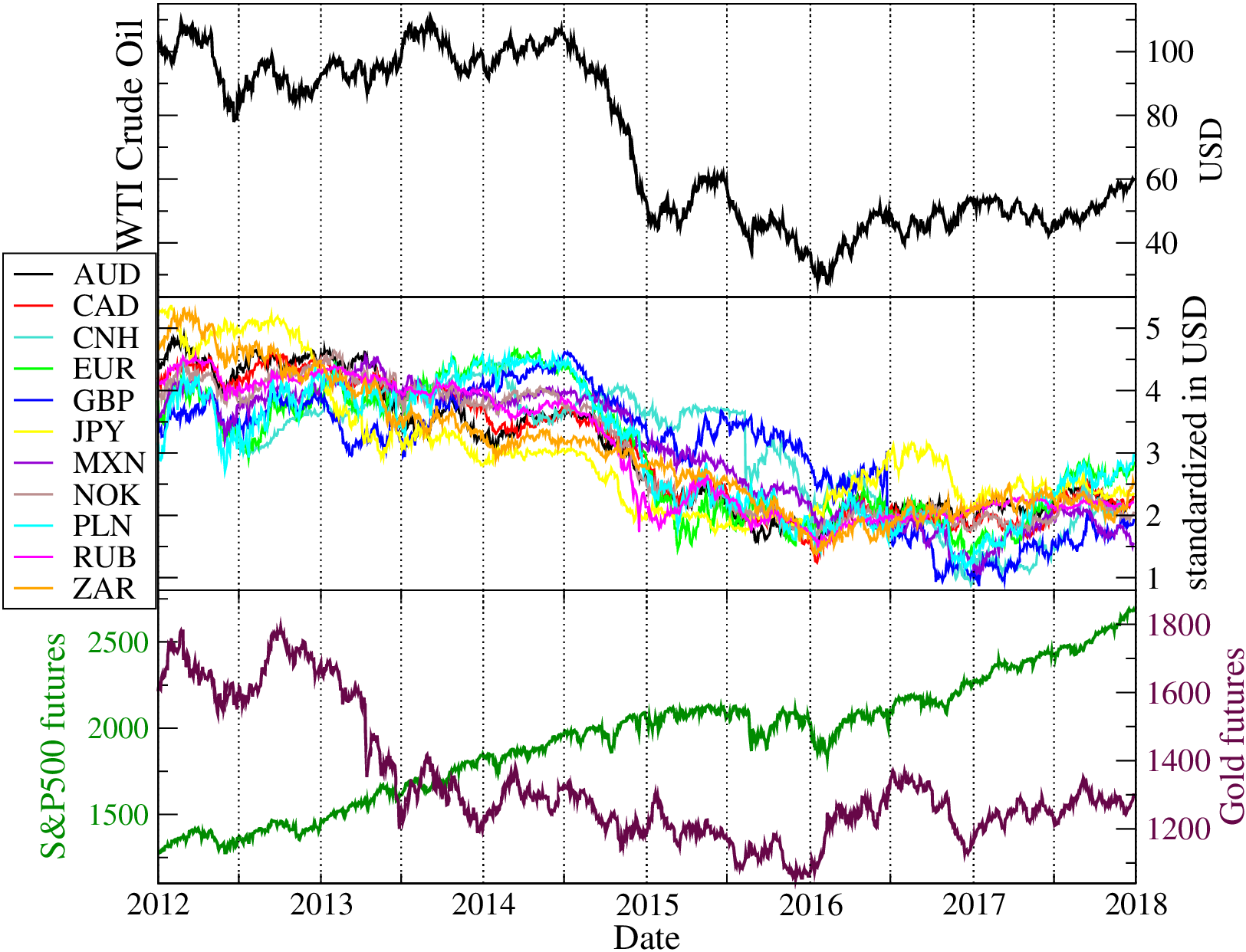}

\caption{(Color online) WTI Crude Oil futures in USD per barrel, gold futures in USD, E-mini S\&P500 futures and standardized currencies in the period Jan 02, 2012 - Dec 29, 2017 but CNH/USD in the period July 02, 2012 - Dec 29, 2017.}
\label{fig:CL}
\end{figure}

The data set used consists of prices of 14 financial instruments comprising high frequency intraday $\Delta t=5$ min quotes in the period between January 02, 2012 and December 29, 2017: (i) WTI Crude Oil futures (CL) - the world’s most liquid and actively traded crude oil contract, traded 23 hours a day, which in addition is known to be well-integrated with the oil spot market in performing the functions of both price discovery and risk transfer \citep{Can2018,SHAO2019}, (ii) E-mini S\&P500 futures (ES) - one of the most efficient, liquid and cost-effective ways to gain market exposure to the S\&P500 Index, traded 23 hours a day, (iii) gold futures (GC) - also traded 23 hours a day and (iv) 10 currencies expressed in US dollar: Australian dollar (AUD/USD), Canadian dollar (CAD/USD), Euro (EUR/USD), British pound (GBP/USD), Japanese yen (JPY/USD), Mexican peso (MXN/USD), Norwegian krone (NOK/USD), Polish zloty (PLN/USD), Russian ruble (RUB/USD) and South African rand (ZAR/USD), as recorded by Swiss forex bank \cite{dukascopy}. This makes this set of data more consistent as it helps to avoid likely mismatches when the data come from different places and sources.
Given the growing economic importance of the Chinese currency it also is included. However, due to different trading properties, People's Bank of China policy changes and data availability in the considered time period it is analysed from July 01, 2012. First of all there are two separate Chinese Renminbi (RMB) markets: onshore Mainland China market - CNY and offshore market - CNH. The first one is controlled by the People's Bank of China (PBOC) and still pegged to the basket of currencies with the current daily trading band 2$\%$ (Historically, on April 16, 2012 - China widens the trading band for the RMB against the dollar to 1$\%$ from 0.5$\%$. Then, it further expanded to 2$\%$ on March 17, 2014. On Aug 2018 the CNY was de-pegged against the USD and referred to a basket of currencies).
Offshore market (CNH) was launched on August 23, 2010. The PBOC cannot directly intervene in the price of CNH, and thus the daily fix and 2$\%$ trading band does not apply to it. As a result, even though some obvious channels for arbitrage do exist \citep{Funke2015}, the pricing gaps between the CNH and CNY markets happen to take place. Since September 2015 the PBOC has thus reportedly taken action in the offshore market when the onshore-offshore gap was increasing \citep{MCCAULEY2018}.
With no controls on capital movements, the CNH market is potentially more affected by external factors which makes it more flexible in links to global markets and thus it potentially is most informative for the present study. Therefore, the offshore Renminbi (CNH/USD), quoted also by the Swiss forex bank Dukascopy from July 2, 2012 is here used. There, in addition, already also exist quantitative studies documenting that the three RMB markets, including the third one existing, offshore RMB non-deliverable forward market (NDF market), largely cointegrate after the reform~\citep{Xu2017,RUAN2019}. Collection of the corresponding charts illustrating price $p(t)$ changes of all the instruments considered in the present study is depicted in Fig. \ref{fig:CL}.

In the corresponding logarithmic returns $r_{\Delta t}=\log(p(t+\Delta t))-\log(p(t))$, where $\Delta t$ stands for the returns' time-lag the weekend gaps, rolling day gaps and the time when some instruments were not traded have been removed thus the series of returns comprise approximately $N$=400000 observations for each of the considered time series.

One straightforwardly accessible and important characteristics of financial time series, of relevance also for the multiscaling analysis, is the functional form of statistical distribution of its fluctuations. Systematic studies \citep{gopi1998,kwapien2012} of the financial return $r_{\Delta t}$ distributions $P_{\Delta t}(r)$ show that in many cases their tails scale according to a power-law $P_{\Delta t}(r) \sim r^{-\gamma}$. For $P_{\Delta t}(r)$ taken in the cumulative form this distribution asymptotically decays according to the inverse cubic power-law, i.e., $\gamma\approx 3$. In the older data coming from the capital market this holds true for $\Delta t$ up to several days but in more recent data \citep{drozdz2003,drozdz2007} $P_{\Delta t}(r)$ bends down sooner (smaller $\Delta t$) towards the normal distribution and the value of the scaling index $\gamma$ becomes thus larger than $3$. This may originate from the acceleration of information flow and a faster disappearance of correlations on larger time scales when going from past to present. 
\begin{figure}[h!]
\centering 
\includegraphics[scale=0.45]{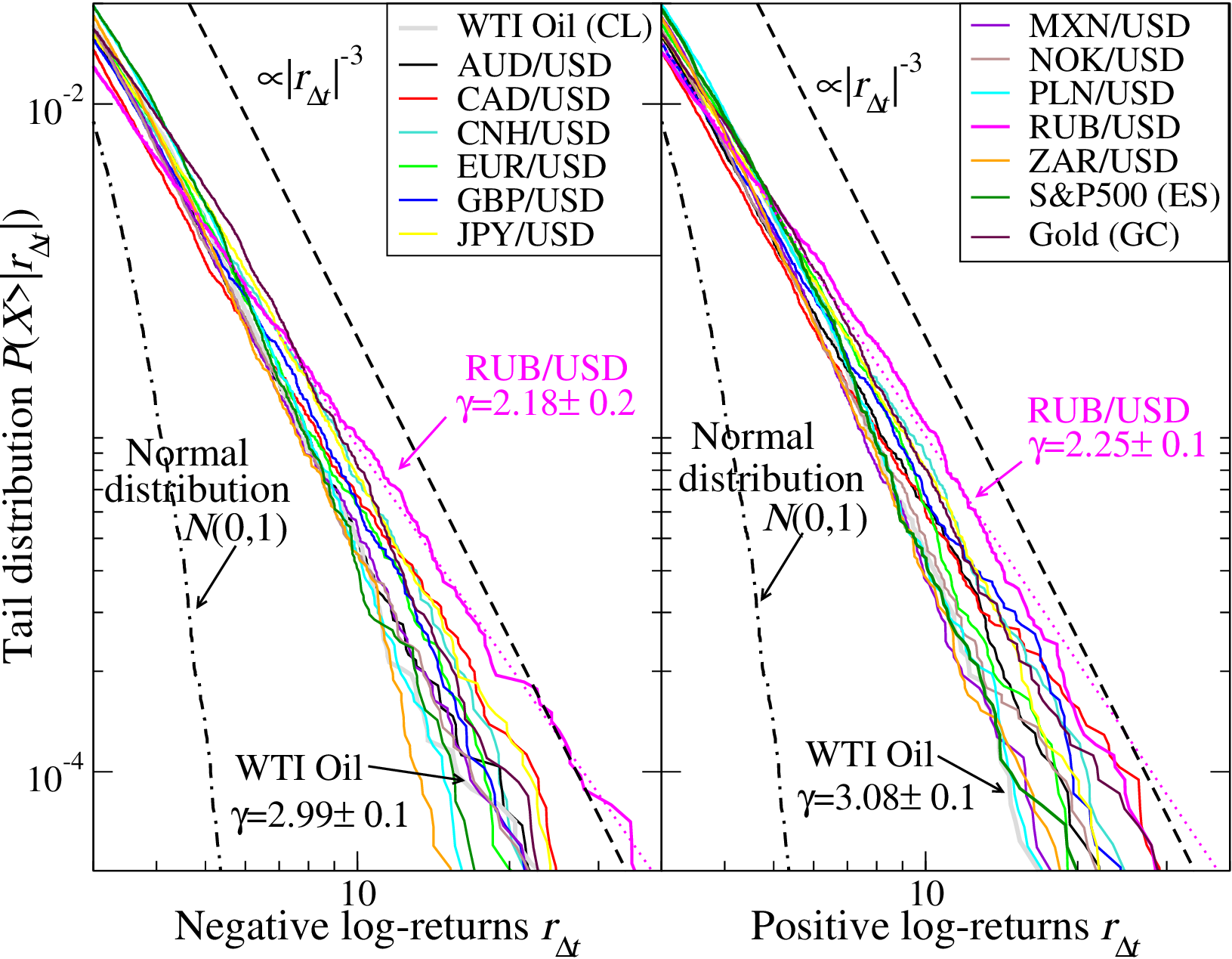}
\caption{(Color online) Log-log plot of the cumulative distributions of normalized  negative (left) and positive (right) returns $r_{\Delta t}(t)$ correspondingly for WTI Crude Oil futures (CL), AUD/USD, CAD/USD, EUR/USD, GBP/USD, USD/JPY, CNH/USD, MXN/USD, NOK/USD, PLN/USD, RUB/USD, ZAR/USD, S\&P500 futures (ES) and gold futures (GC). The dashed line represents the inverse cubic power-law and the dotted lines a straight line fits for RUB/USD with the corresponding $\gamma = 2.18 \pm 0.2$  and $\gamma = 2.25 \pm 0.1$.}
\label{fig:distribution}
\end{figure}
Fig. \ref{fig:distribution} shows the cumulative distributions of the normalized returns  $r_{\Delta t}(t)$, separately for their negative and positive values, obtained for each of the time series specified above for $\Delta t = 5$ min. The two wings of this distribution to a good approximation are symmetric.

Furthermore, for such a short time-lag $\Delta t$ the distributions in this figure appear largely consistent with the inverse cubic power-law. The most outlying distribution whose tail is significantly thicker is the one that corresponds to the Russian ruble with $\gamma \approx 2.2$.

\section{Multifractal formalism}
\label{metdesc}

Research here conducted and addressing the multifractality aspects of time series is based on the formalism of multifractal cross-correlation analysis (MFCCA) method as introduced by \cite{oswiecimka2014}. This method represents a consistent generalization of the detrended cross-correlation approach (DCCA) proposed by \cite{podobnik2008} and extended by \cite{zhou2008}. The corresponding MFCCA methodology allows to quantitatively characterize the scaling properties of single time series as well as a degree of the multifractal cross-correlation between any two times series. This methodology also allows to introduce the $q$-dependent detrended cross-correlation coefficient $\rho_q$, as proposed by \cite{kwapien2015}, which filters out the strength of cross-correlations varying with the size of fluctuations when such effects take place.  

One thus considers a pair of time series ${x(i)}_{i=1,...,T}$ and ${y(i)}_{i=1,...,T}$ divided into $2 M_s$ separate boxes of length $s$ (i.e., $M_s$ boxes starting from the opposite ends). Then, the detrending procedure is applied by calculating in each box $\nu$ ($\nu=0,...,2 M_s - 1$) the residual signals $X,Y$ equal to the difference between the integrated signals and the $m$th-order polynomials $P^{(m)}$ fitted to these signals:

\begin{eqnarray}
X_{\nu}(s,i) = \sum_{j=1}^i x(\nu s + j) - P_{X,s,\nu}^{(m)}(j),\\
Y_{\nu}(s,i) = \sum_{j=1}^i y(\nu s + j) - P_{Y,s,\nu}^{(m)}(j).
\end{eqnarray}
In typical cases an optimal choice corresponds to $m=2$ \citep{oswiecimka2006,oswiecimka2013} and such is used in the present work. Next, the covariance and the variance of $X$ and $Y$ in a box $\nu$ is calculated according to the definition:
\begin{eqnarray}
\label{eq::covariance}
f_{XY}^2(s,\nu) = {1 \over s} \sum_{i=1}^s X_{\nu}(s,i) Y_{\nu}(s,i),\\
f_{ZZ}^2(s,\nu) = {1 \over s} \sum_{i=1}^s Z_{\nu}^2(s,i),
\label{eq::variance}
\end{eqnarray}
where $Z$ means either $X$ or $Y$. These quantities can be used to define a family of the fluctuation functions of order $q$ \citep{oswiecimka2014}:

\begin{eqnarray}
\label{eq::covariance.q}
F_{XY}^q(s) = {1 \over 2 M_s} \sum_{\nu=0}^{2 M_s - 1} {\rm sign} \left[f_{XY}^2(s,\nu)\right] |f_{XY}^2(s,\nu)|^{q/2},\\
F_{ZZ}^q(s) = {1 \over 2 M_s} \sum_{\nu=0}^{2 M_s - 1} \left[f_{ZZ}^2(s,\nu)\right]^{q/2}.
\label{eq::variance.q}
\end{eqnarray}
The real parameter $q$ plays the role of a filter as it amplifies or suppresses the intra-box variances and covariances in such a way that for large positive $q$-values only the boxes (of size $s$) with the highest fluctuations contribute substantially to the sums in Eqs. (\ref{eq::covariance.q} and \ref{eq::variance.q}) while for small negative $q$-values only the boxes with the smallest fluctuations provide a dominant contribution. 

Multifractal cross-correlation is expected to manifest itself in a power-law dependence of $F_{XY}^{q}(s)$ and thus in the following relation:
\begin{equation}
F_{XY}^{q}(s)^{1/q}=F_{XY}(q,s) \sim s^{\lambda(q)},
\label{Fxy}
\end{equation}
where $\lambda(q)$ is an exponent that quantitatively characterizes the fractal aspects of cross-correlations. For the monofractal cross-correlation the exponent $\lambda(q)$ is $q$-independent. A $q$-dependence of $\lambda(q)$ signals the multifractal character of cross-correlations under study.

$F_{ZZ}$ characterizes the scaling properties of single time series:
\begin{equation}
F_{ZZ}^{q}(s)^{1/q}=F_{ZZ}(q,s) \sim s^{h(q)},
\label{Fzz}
\end{equation}
and $h(q)$ stands for the generalized Hurst exponent. For the monofractal time series $h(q)=const$ whereas for the multifractal signals $h(q)$ is a decreasing function of $q$. The corresponding singularity spectrum $f(\alpha)$ can be calculated according to the following relations \citep{kantelhardt02}:

\begin{equation}
\tau(q)=qh(q)-1,
\end{equation}
\begin{equation}
\alpha=\tau'(q)\quad \textrm{and} \quad f(\alpha)=q\alpha-\tau(q),
\end{equation}
where $\alpha$ is called the singularity exponent or the H\"older exponent and $f(\alpha)$ is the corresponding singularity spectrum often referred to as the multifractal spectrum. This special case of single time series corresponds to the celebrated Multifractal Detrended Fluctuation Analysis (MFDFA) of \cite{kantelhardt02}.  

For time series generated by the model mathematical cascades the singularity spectrum $f(\alpha)$ corresponding to the moments (Eq.~{\ref{eq::variance.q}) of order ranging between $-q$ and $+q$ assumes form of a symmetric upper most fragment of an inverted parabola. As shown by \cite{drozdz2015} the realistic time series are often distorted in their hierarchical organization as compared to a purely uniform organization of mathematical cascades. Such effects manifest themselves in asymmetry of $f(\alpha)$ and they also may be crucially informative for identifying the composition of time series. One possibility to globally characterize this kind of asymmetry of $f(\alpha)$ is through the asymmetry parameter \citep{drozdz2015}:
\begin{equation}
A_{\alpha} = ({\Delta \alpha}_L - {\Delta \alpha}_R) / ({\Delta \alpha}_L + {\Delta \alpha}_R),
\label{ap}
\end{equation}
where ${\Delta \alpha}_L = {\alpha}_0 - {\alpha}_{min}$ and ${\Delta \alpha}_R = {\alpha}_{max} - {\alpha}_0$ and $\alpha_{min}$, $\alpha_{max}$, $\alpha_0$ denote the beginning and the end of $f(\alpha)$ support, and the $\alpha$ value at maximum of $f(\alpha)$ (which corresponds to $q=0$), respectively. The positive value of $A _\alpha$ reflects the left-sided asymmetry of $f(\alpha)$, i.e. its left arm is stretched with respect to the right one. This indicates a more developed multifractality on the level of large fluctuations in the time series. Negative $A _\alpha$, on the other hand, reflects the right-sided asymmetry of the spectrum and thus indicates the smaller fluctuations as a dominant multifractality source. 

A family of the fluctuation functions as defined by Eq.~(\ref{Fxy}) can also be used to define a $q$-dependent detrended cross-correlation ($q$DCCA) \citep{kwapien2015} coefficient:
\begin{equation}
\rho_q(s) = {F_{XY}^q(s) \over \sqrt{ F_{XX}^q(s) F_{YY}^q(s) }},
\label{rhoq}
\end{equation}
which allows to quantify the degree of cross-correlations between the two time series $x(i)$, $y(i)$ after detrending and at varying time scales $s$. By varying the parameter $q$ one can identify the range of detrended fluctuation amplitudes for which the two signals $x(i)$ and $y(i)$ are correlated most. This filtering ability of $\rho_q(s)$ constitutes an important advantage over the more conventional methods since cross-correlations among the realistic time series are usually not uniformly distributed over their fluctuations of different magnitude \citep{kwapien2017}. Of course, the $\rho_q(s)$ coefficient can be used to quantify cross-correlations also between series that develop no well established multifractality characteristics.

\section{Multifractality of individual series}
\label{MFDFA1}
A necessary, thought not sufficient, condition for the multifractal cross-correlations to occur among price changes of different financial assets is multifractality of each in separation. The first step towards such multifractal analyses is to calculate the fluctuation functions according to Eq. (\ref{Fzz}), for each individual time series. In order to eliminate a possible bias in estimating the fluctuation functions the parameter $q$-values are taken from the interval $q\in [-4,4]$ which, due to the inverse cubic power-law \citep{gopi1998,drozdz2003} governing asymptotic distribution of large returns also in the present case as shown in Section \ref{data}, prevents entering the region of divergent moments for $q>4$. The minimum and maximum scales depend on the length $N$ of the time series under study. In the present study they are set as $s=10$ (50min) and $s=3750$ (14 trading days), correspondingly. One thus obtains a family of $F(q,s)$ functions displayed in Fig.~\ref{fig:qvariance}. They are seen to follow a very convincing power-law behaviour that can be characterized by sets of non-uniform scaling exponents $h(q)$ whose $q$-dependence is shown in the corresponding insets. 

\begin{figure}[h!]
\centering 
\includegraphics[scale=0.45]{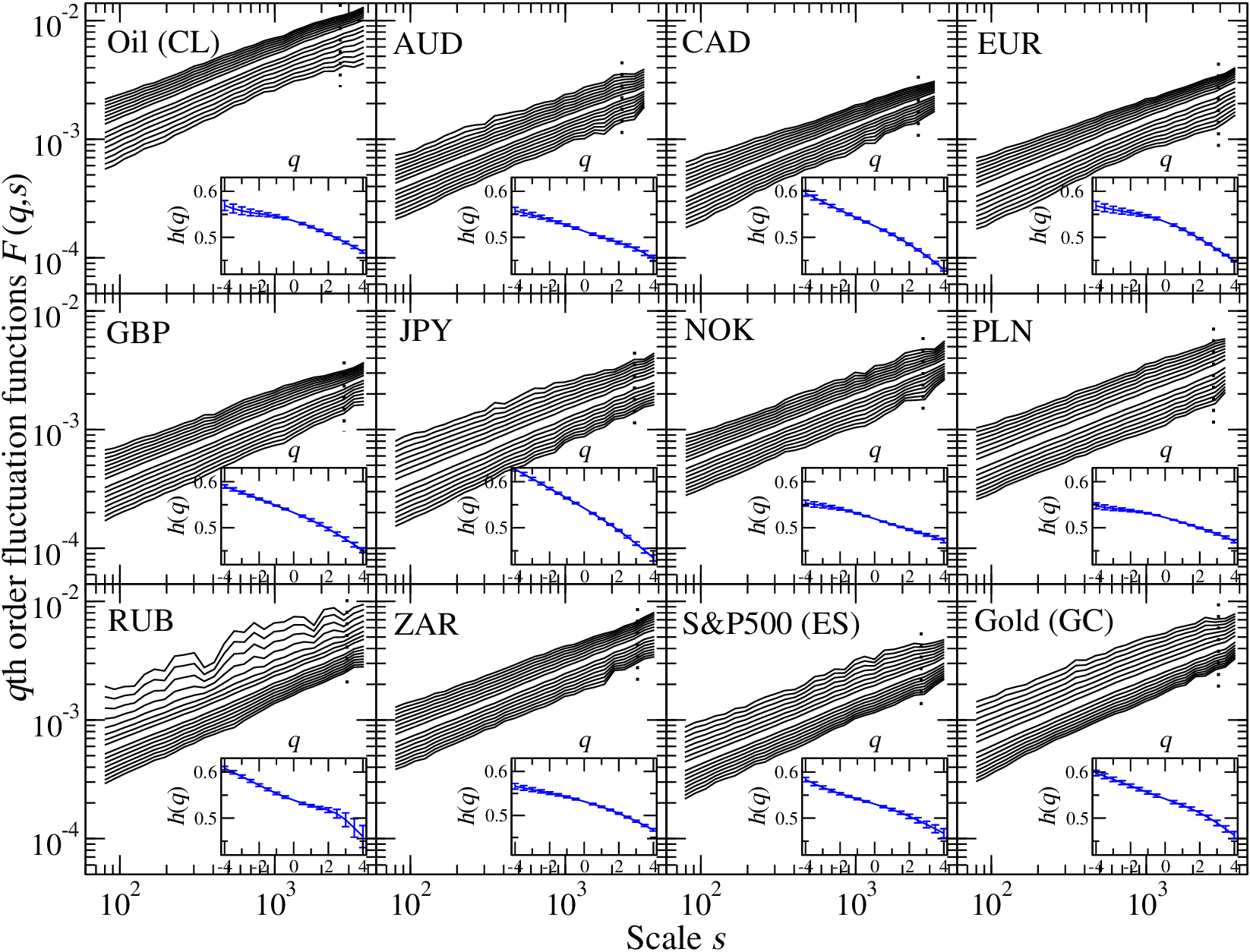}

\caption{Family of the $q$th-order fluctuation functions $F(q,s)$ for different values of $q\in [-4,4]$ with step 0.5 (the topmost one represents $q=4$) calculated for WTI Crude Oil futures (CL), AUD/USD, CAD/USD, EUR/USD, GBP/USD, JPY/USD, NOK/USD, PLN/USD, RUB/USD, ZAR/USD, S\&P500 futures (ES) and gold futures (GC). Insets illustrate the corresponding generalized Hurst exponents $h(q)$.}
\label{fig:qvariance}
\end{figure}

This multiscaling behaviour has its reflection in the shape of the multifractal spectra $f(\alpha)$ depicted in Fig.\ref{fig:singspektrum}. In consequence, each of the considered time series can be regarded as having a multifractal organization with a well-developed multifractal spectrum which, characteristically, are left-sided $(A_{\alpha}>0)$. Such effects of asymmetry of the multifractal spectra $f(\alpha)$, either left- or right-sided, indicate non-uniformity \citep{drozdz2015} in the hierarchical organization of the time series. The left-side of $f(\alpha)$ is projected out by the positive $q$-values thus it reflects organization of the large fluctuations while the negative $q$-values determine the right-side of $f(\alpha)$. Consequently, the left-sided asymmetry of $f(\alpha)$ indicates a more pronounced multifractality on the level of large fluctuations while the opposite applies to the right-side. As Fig.~\ref{fig:singspektrum} shows, the left-sided asymmetry is much more common among the assets here considered which indicates that they develop a more pronounced multifractality on the level of large fluctuations and that their smaller fluctuations are more noisy. 

\begin{figure}[h!]
\centering 
\includegraphics[scale=0.45]{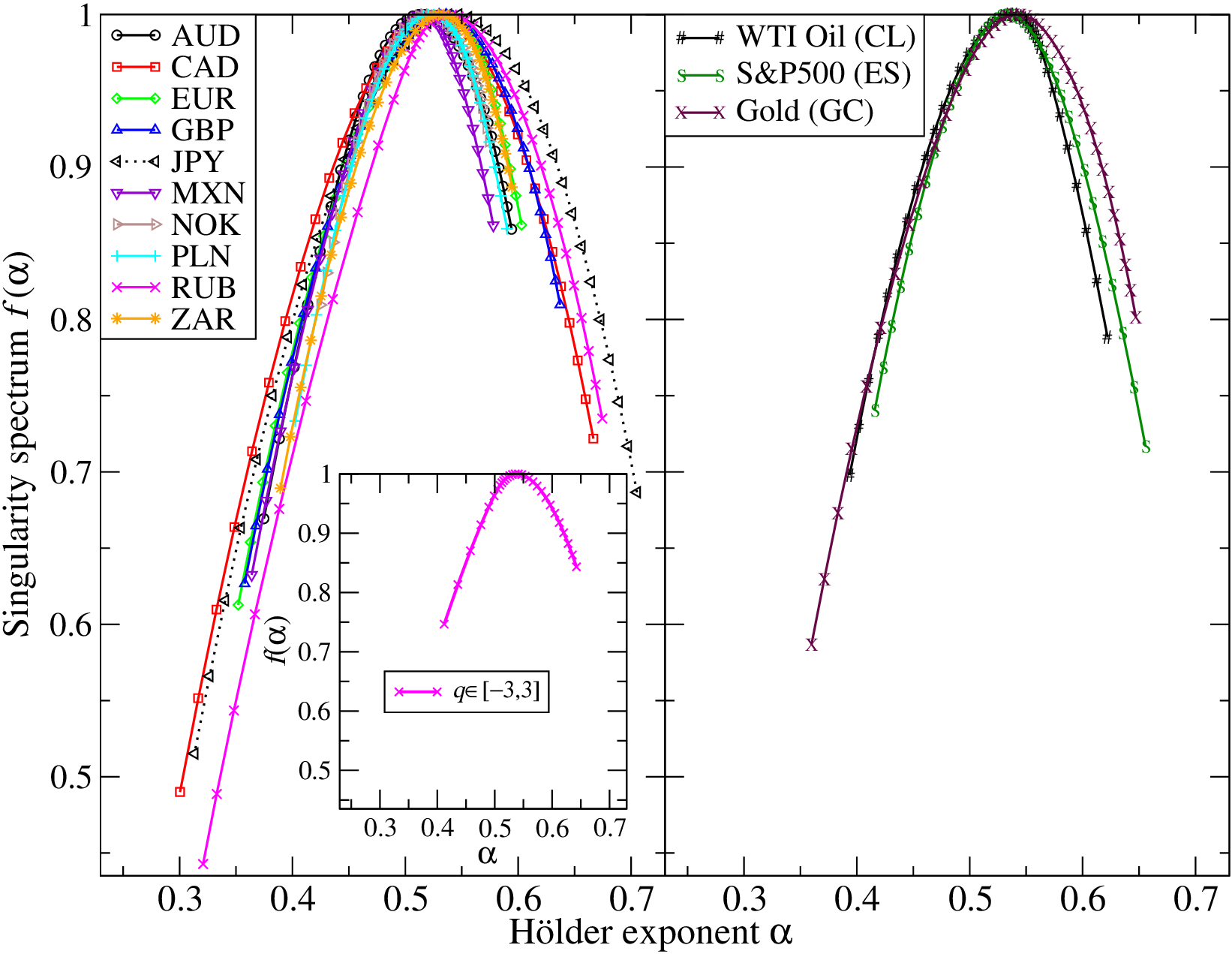}

\caption{(Color online) Singularity spectra $f(\alpha)$ calculated for all the considered time series using $q\in [-4,4]$. Inset shows $f(\alpha)$ of the Russian ruble for $q\in [-3,3]$.} 
\label{fig:singspektrum}
\end{figure}

The broadest multifractal spectra, with a sizeable left-sided asymmetry at the same time, are observed for RUB/USD ($\Delta \alpha \approx 0.37$ and $A_{\alpha} \approx 0.24$) and JPY/USD exchange rate ($\Delta \alpha \approx 0.39$ and $A_{\alpha} \approx 0.17$) whereas the narrowest spectrum is obtained for NOK/USD with $\Delta \alpha \approx 0.17$ but with an even larger relative asymmetry as expressed by $A_{\alpha} \approx 0.34$. In fact it is the ruble which develops the longest left wing in its $f(\alpha)$ but its evaluation needs some care, however. As it is seen in Fig.~\ref{fig:qvariance} it only is this case that the fluctuation functions $F(q,s)$ for larger positive values of the $q$-parameter are somewhat perturbed along their power-law trend. This may partly have to do with the fact that the ruble distribution of return fluctuations develops thicker tails in Fig.~\ref{fig:distribution} than all the remaining instruments shown there and thus the moments of order $q \ge 3$ are divergent in the limit of an infinite series. Restricting the procedure to in this particular case 'legal' interval of $q\in [-3,3]$ results in the spectrum displayed in the inset of Fig.~\ref{fig:singspektrum}. What is natural, within such an interval of $q$-values the ruble's multifractal spectrum $f(\alpha)$ is seen narrower but at the same time its asymmetry gets reduced to $A_{\alpha} \approx 0.10$. Overall, however, as Fig.~\ref{fig:singspektrum} shows, all the instruments here studied, the currencies as well as the oil, gold and S\&P500, develop pronounced multifractal spectra and the left-sided asymmetry is basically common. As far as this latter characteristics is concerned the S\&P500 constitutes an exception. Its $f(\alpha)$ is close to a symmetric shape with an even some tendency to right-sidedness ($A_{\alpha} \approx -0.09$). 

The related multifractal characteristics of the Chinese currency CNH/USD, due to its somewhat different rules of trading over the period here considered and a half year shorter interval of data availability, are shown in Section \ref{CNH} together with the other quantitative characteristics of its correlation to oil.

\section{Multifractality aspects of cross-correlations}
\label{MFCCAa}

In order to extend this analysis towards the main issue of the present study, i.e., quantification of cross-correlations between the WTI Crude Oil futures (CL) and other financial instruments here considered the corresponding $q$th order fluctuation functions $F_{XY}(q,s)$ according to Eq.~(\ref{Fxy}) are calculated and this is followed by the search for some possible manifestations of scaling. 
Indeed, the results of such calculations shown in Fig.~\ref{fig:FqCL} reveal that quite a convincing scaling behaviour of $F_{XY}(q,s)$ can be observed in all the cases considered. This indicates that there is some synchrony in evolution of the corresponding assets even on the level of their multifractal organization. The power-law behaviour of $F_{XY}(q,s)$ is, however, seen exclusively for the positive $q$-values and therefore these are shown in Fig.~\ref{fig:FqCL} with the lower limiting values of $q$ listed explicitly. Below those values of the cross-correlation fluctuation functions $F_{XY}(q,s)$ start fluctuating irregularly, occasionally assuming even the negative values, similarly as in certain other financial cases studied before \citep{Rak2015}. Such qualitative transitions - scaling versus non-scaling - in the behaviour of $F_{XY}(q,s)$ are in fact compatible with the left-sidedness of the multifractal spectra of single time series in Fig.~\ref{fig:singspektrum}. The multifractal cross-correlations posses more preferential conditions to take place on the level of larger fluctuations because they develop richer multifractality and these are seen primarily through the positive parameters $q$. Out of all the assets here considered the Japanese yen is seen to be multifractally correlated least with the oil as both the quality of scaling of the corresponding $F_{XY}(q,s)$ is worst and, in addition, it terminates already at around $q=0.8$.

\begin{figure}[h!]
\centering 
\includegraphics[scale=0.45]{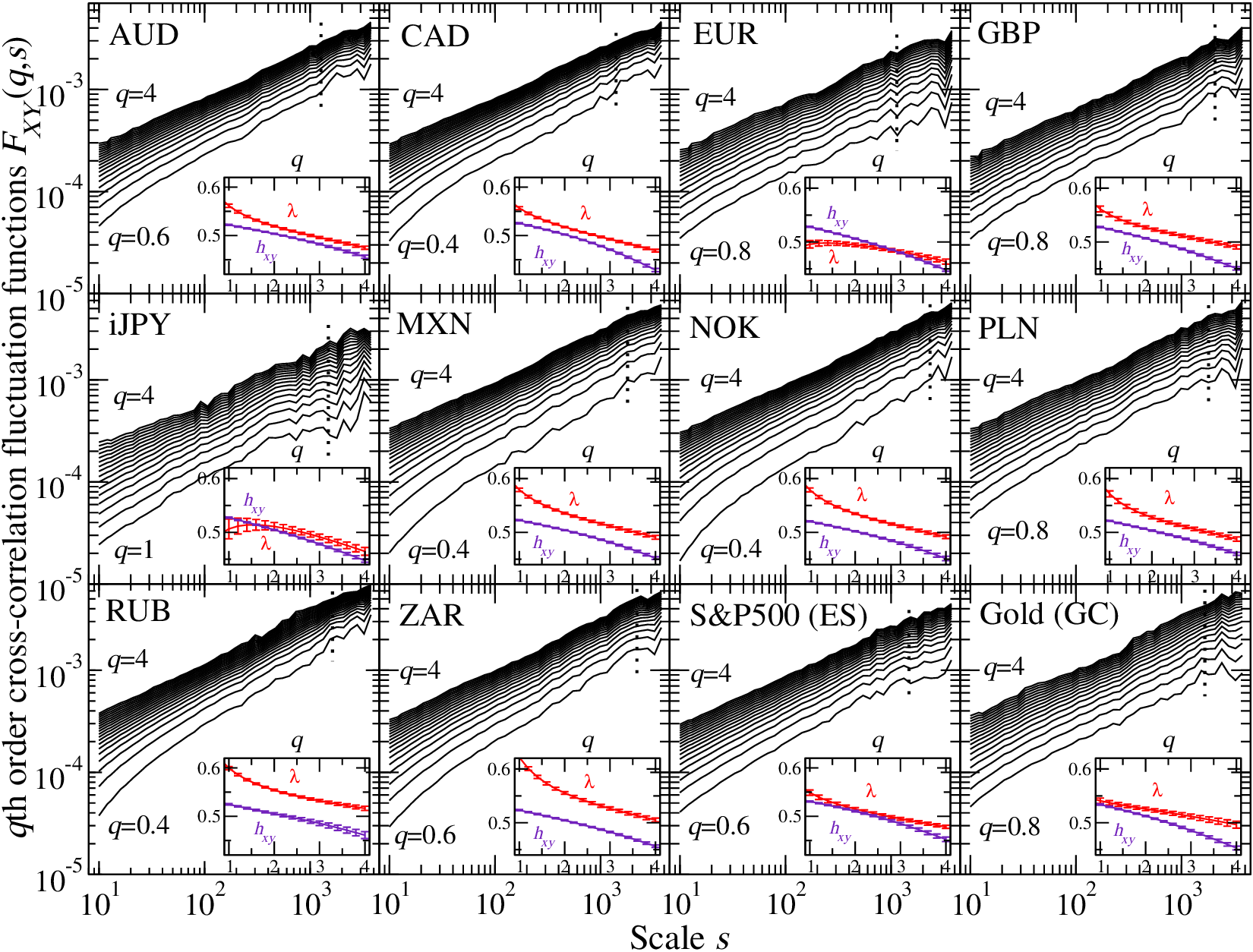}

\caption{Family of the $q$th-order fluctuation cross-covariance functions $F_{XY}(q,s)$ for different values of $q$ with the step of 0.2 calculated for WTI Crude Oil futures (CL) versus AUD/USD, CAD/USD, EUR/USD, GBP/USD, USD/JPY, MXN/USD, NOK/USD, PLN/USD, RUB/USD, ZAR/USD, S\&P500 futures (ES) and gold futures (GC). The topmost lines correspond to $q=4$ and the lowest ones to those $F_{XY}(q,s)$ that still develop a relatively regular shape with the corresponding limiting $q$-value explicitly listed. Insets illustrate the resulting $q$-dependence of $\lambda(q)$ versus the average of the generalized Hurst exponents $h_{xy}(q) = (h_x(q)+h_y(q))/2$ of the two series $x(i)$ and $y(i)$ under study.}
\label{fig:FqCL}
\end{figure}

Even though the fluctuation functions $F_{XY}(q,s)$ quantifying the multifractality aspects of cross-correlations between the oil price changes and several world currency exchange rates, as displayed in Fig.~\ref{fig:FqCL}, look largely similar, some more subtle effects can be detected by comparing the corresponding scaling exponents $\lambda(q)$ and the average of generalized Hurst exponents: 
\begin{equation}
h_{xy}(q) = (h_x(q)+h_y(q))/2.
\label{hqxy}
\end{equation}
While $h_{xy}(q)$ behaves alike in all the cases considered a variety of the $q$-dependences of $\lambda(q)$ can be observed. As it is shown in Fig.~\ref{fig:lambda_hsred_CL} this results in correspondingly different $q$-dependences of:
\begin{equation}
d_{xy}(q) = \lambda(q) - h_{xy}(q),
\label{dqxy}
\end{equation}
and reflects a different speed of the covariance accumulation in Eq.(\ref {eq::covariance.q}) with an increasing scale $s$. In majority of the cases this increase is faster than the one of its counterpart variances in the two series and therefore $\lambda(q)$ is usually larger than $h_{xy}(q)$. In some rarer cases, like CL vs. JPY or CL vs. EUR, the opposite, however, applies for the smaller $q$-values. This signals that certain specific elements of the dynamics of just these two currencies are distinct. Indeed, as it can be verified by an explicit calculation, their multifractal cross-correlation with the other currencies appears to be weaker than what is typical among currencies. As an example, there is essentially no this kind of cross-correlation between JPY and MXN, i.e., no graph analogous to those of Fig.~\ref{fig:FqCL} can be drawn. A more systematic study of the related effects may be very instructive but it is however beyond the scope of the present oil dedicated contribution. 

\begin{figure}[h!]
\centering 
\includegraphics[scale=0.45]{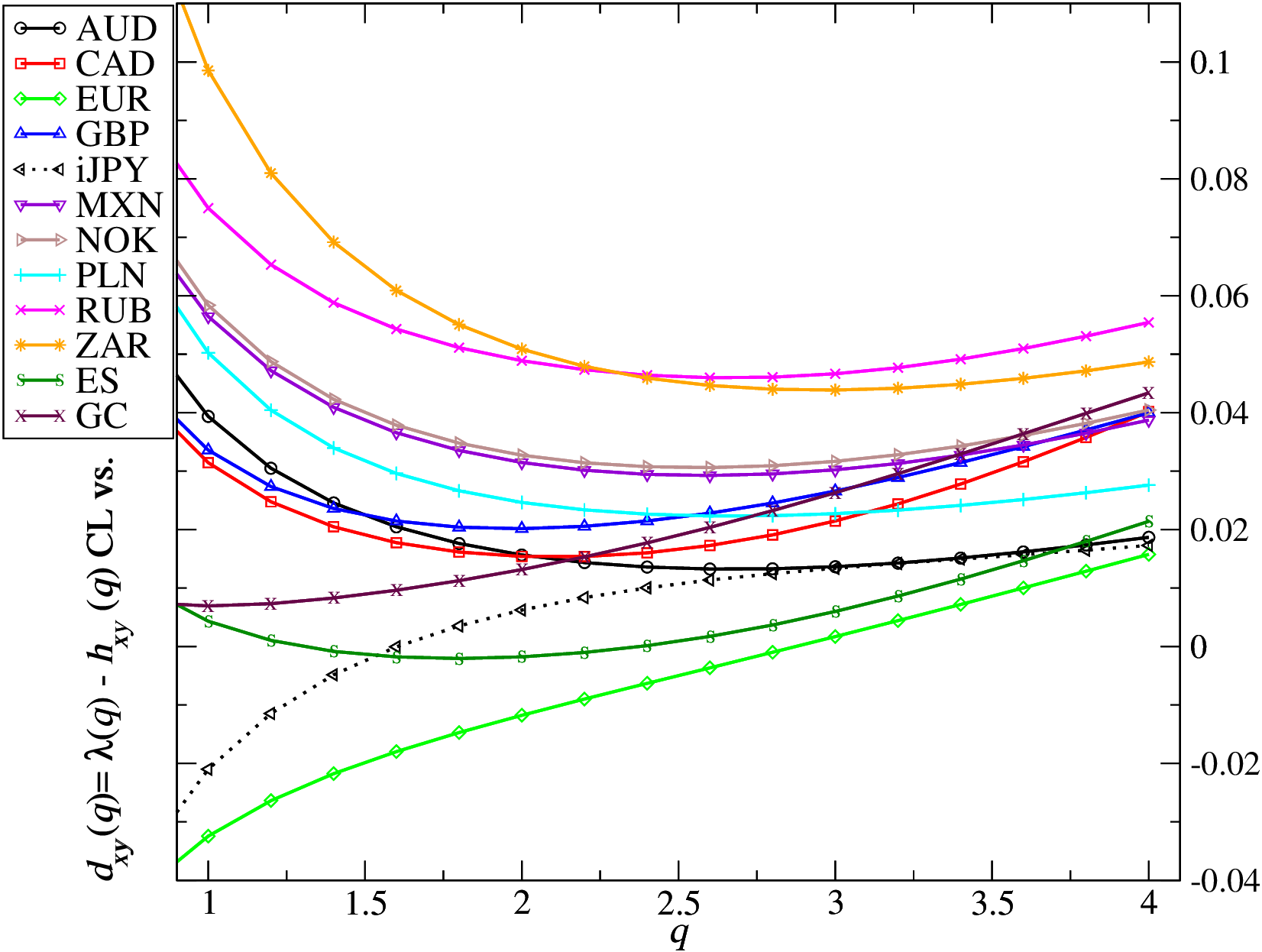}

\caption{(Color online) Differences between multifractal cross-correlation scaling exponents $\lambda(q)$ and the average generalized Hurst exponents $h_{xy}(q)$ estimated for $q\in [1,4]$.} 
\label{fig:lambda_hsred_CL}
\end{figure}

\section{Cross-correlations quantified using $\rho_q$ coefficient}
\label{Pq}

A complementary measure of cross-correlations is based on the $q$-dependent detrended cross-correlation coefficient $\rho_q$ calculated according to the formula of Eq.~(\ref{rhoq}). The time scale $s$-dependence of such coefficients between WTI Crude Oil futures and other financial instruments here considered for $q=1,2,3$ and $4$ is shown in Fig.~\ref{fig:pDCCACL}. Except for the Japanese yen (JPY/USD), which in order to invert it into the positive values in this figure is represented by its inverse (iJPY = USD/JPY), all the currencies (expressed in USD), the S\&P500 futures (ES) and gold futures (GC) are seen to be positively correlated with the oil. 

The strongest correlation is observed for the Russian ruble (RUB/USD) and for the Canadian dollar (CAD/USD). These correlations, as well as those of the remaining currencies, systematically decrease with the increasing  $q$-value by on average about a factor of 3 when $\rho_1$ and $\rho_4$ are compared. This indicates that they take place primarily on the level of smaller and medium size fluctuations. In this $q$-dependence of $\rho_q$ some reordering of magnitudes can also be observed. Out of the two most with the oil correlated currencies for $q=1$ it is the Russian ruble that dominates and the Canadian dollar is the second while for $q=4$ the latter overtakes. Similar reordering one can observe for the other less with oil correlated currencies as well. Interestingly, the Euro and British pound belong to those currencies that are weakly correlated with oil. The least in magnitude correlated, systematically, for all the $q$-values, remains the Japanese yen and this correlation, as pointed out above, is negative in sign. The negative correlation of the Japanese yen to oil may originate from the facts that Japan is one of the world leading oil net importers \citep{Lizardo2010} and that it typically has the negative interest rates which favours activation of its carry trade effect in relation to oil \citep{Lu2017}. Statistical significance of all these results is tested against the null hypothesis of randomly shuffled original series. The result representing average over 100 realizations of such surrogates for each original time series is depicted by the dotted line in Fig.~\ref{fig:pDCCACL}. Clearly, all the original results stay very convincingly above this line. 

\begin{figure}[h!]
\centering 
\includegraphics[scale=0.45]{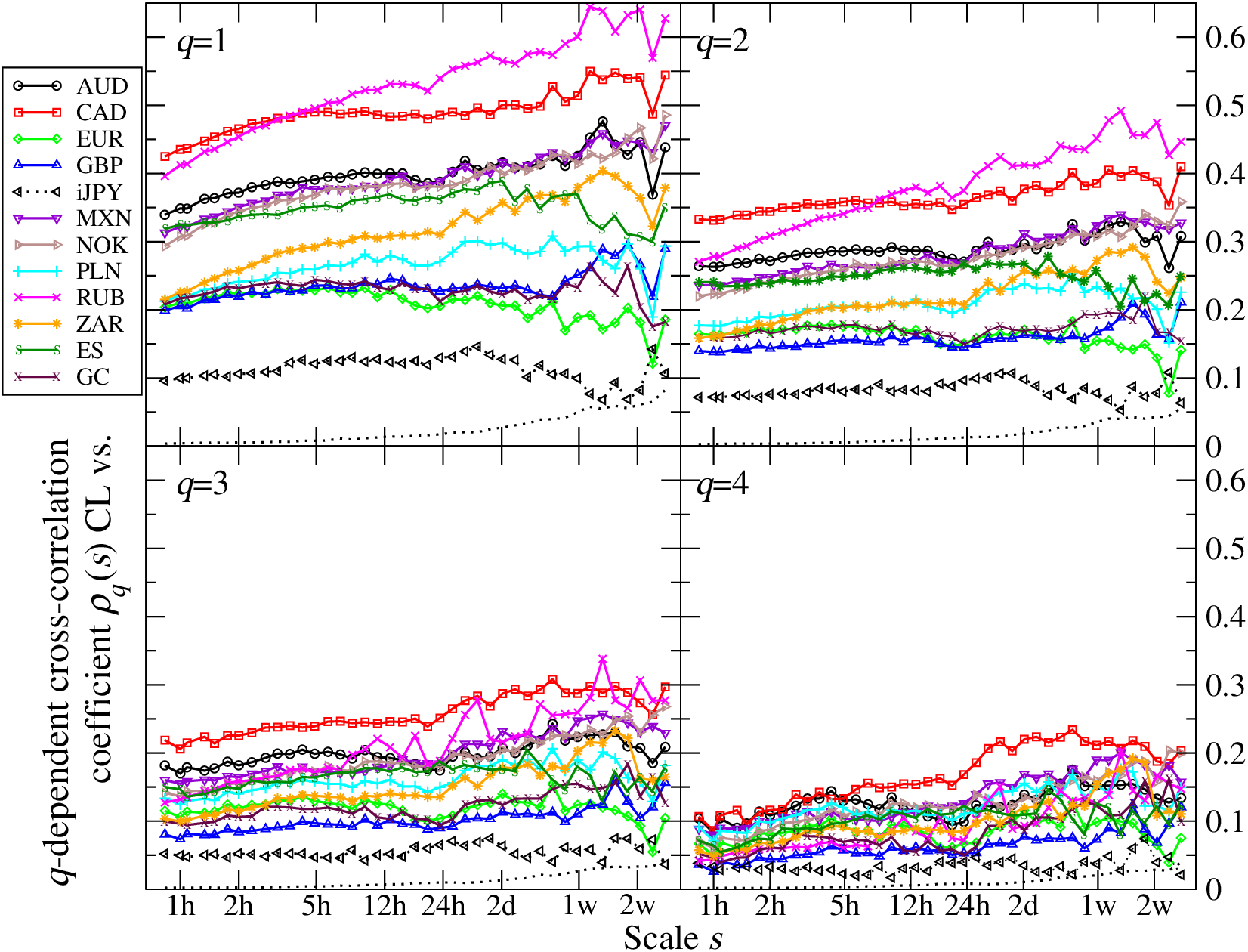}
\caption{(Color online) The $q$-dependent detrended cross-correlation coefficient $\rho_q$ between WTI Crude Oil futures (CL) and other financial instruments of Fig.~\ref{fig:FqCL} as a function of the temporal scale $s$ for $q=1$, $q=2$, $q=3$ and $q=4$,  as well as the standard deviation $\sigma_{\rho}(q,s)$ obtained from 100 independent realizations of the shuffled surrogate data (dotted grey lines).}
\label{fig:pDCCACL}
\end{figure}

The strongest cross-correlations detected for all the instruments here studied versus CL for $q = 1$, thus for the smaller amplitude fluctuations, and the weakest cross-correlations associated with the largest considered fluctuations ($q = 4$) parallel the results obtained for the mature stock markets and for the Forex by \cite{kwapien2015} and by \cite{Zhao2017}.
 
These present results show that generally currencies of the countries selling oil or other natural resources, like Australian dollar (AUD/USD) (also coal producer), Canadian dollar (CAD/USD), Mexican peso (MXN/USD), Norwegian krone (NOK/USD)),  Russian ruble (RUB/USD) and South African rand (ZAR/USD) - the so called commodity currencies - develop stronger correlations with the oil price changes. Weaker correlations reveal currencies of the countries with looser connections to the oil market or to the oil importers. These include Euro (EUR/USD), British Pound (GBP/USD), Japanese yen (JPY/USD) or Polish zloty (PLN/USD).
The stronger correlation between CL and S\&P500 may originate from the substantial contribution of the oil companies in the index (such as Exxon and Chevron). Other relevant factor may be related to the so-called risk on/off effect \citep{riskonoffm}. Both instruments are risky assets which can make them co-move \citep{CLSP500}.

\subsection{Scale dependence of $\rho_q$}

One further significant and characteristic feature of the $\rho_q(s)$ coefficients in Fig.~\ref{fig:pDCCACL} is their scale $s$-dependence. Various forms of scale-dependences in cross-correlations, somewhat complementary to the widely known Epps effect \citep{Epps,kwapien2004,Toth2009,drozdzepps} which refers to an increase of correlations with the  decreasing frequency of probing price changes, already appear in the literature, see e.g. \cite{Ma2013,Kristo2014,Ma2014,Pal2014,Reboredo2014,kwapien2015,Barunik2016,Li2016,Yang2016,Hussain2017,Zhao2017,Ferreira2019} with no systematic analysis of its origin and meaning. Here, however, quite a systematic correspondence between the scale $s$-dependence of $\rho_q(s)$ in Fig.~\ref{fig:pDCCACL} and $d_{xy}(q)$ displayed in Fig.~\ref{fig:lambda_hsred_CL} can be traced. The larger is $d_{xy}(q)$ the faster on average is increase of $\rho_q(s)$ with $s$ increasing. Even more, comparing the four panels of Fig.~\ref{fig:pDCCACL} which illustrate the $q$-dependence of these effects one also finds consistency. Smaller values of $d_{xy}(q)$ are associated with the weaker $s$-dependences of $\rho_q(s)$ and even the negative value of $d_{xy}(q)$, like in the Euro case at $q=1$, is seen to correspond to a decreasing $\rho_q(s)$. 

According to Eq.~(\ref{rhoq}) for the multifractal series such relations are in fact natural. The positive $d_{xy}(q)$ means that $\lambda(q) > (h_x(q)+h_y(q))/2$ and thus the numerator in this Equation increases faster with the scale $s$ than the denominator. The opposite applies for the negative $d_{xy}(q)$. In more rigorous terms the power-law relations imply $F_{XY}(q,s) \sim s^{\lambda(q)}$ and  $\sqrt{F_{XX}(q,s)F_{YY}(q,s)} \sim s^{\frac{h_x(q)+h_y(q)}{2}} = s^{h_{xy}(q)}$. By introducing the corresponding proportionality coefficients this can be written as $F_{XY} (q,s)=a_{xy}s^{\lambda(q)}$, $F_{XX}(q,s)=a_xs^{h_x(q)}$ and $F_{YY}(q,s)=a_ys^{h_y(q)}$. From the Cauchy-Schwartz inequality if follows \citep{kwapien2015} that: 
\begin{equation}
 F_{XY}^q(s) \le \sqrt{ F_{XX}^q(s) F_{YY}^q(s) }, \quad q \ge 0.
\label{eq::cauchy-schwarz-like}
\end{equation}
By substitution one thus receives:
\begin{equation}
(a_{xy})^qs^{\lambda(q)\cdot q}\leq (a_xa_y)^{q/2}s^{(h_x(q)+h_y(q))\cdot q/2}.
\label{eq::ca}
\end{equation}
For $q>0$ this leads to:
\begin{equation}
\lambda(q) \leq \log_s(\frac{\sqrt{a_xa_y}}{a_{xy}}) + \frac{h_x(q)+h_y(q)}{2}.
\label{eq:ca1}
\end{equation}
For $q>0$ the proportionality coefficients, responsible for the relative placement of functions: $F_{XY}^q(s) $ and $\sqrt{ F_{XX}^q(s) F_{YY}^q(s)}$ must obey the inequality $a_{xy} \leq \sqrt{a_xa_y} $ therefore $\log_s(\frac{\sqrt{a_xa_y}}{a_{xy}})$ is positive thus a difference between $\lambda(q)$ and $h_{xy}(q)$ in either direction is allowed. This is under assumption that the time series is finite. For $s \to \infty$ only the case of $\lambda(q) \leq  h_{xy}(q)$ is allowed \citep{He2011,Kristo2015}.

\subsection{Characteristics of Chinese currency in relation to oil}
\label{CNH}
\begin{figure}[h!]
\centering 
\includegraphics[scale=0.45]{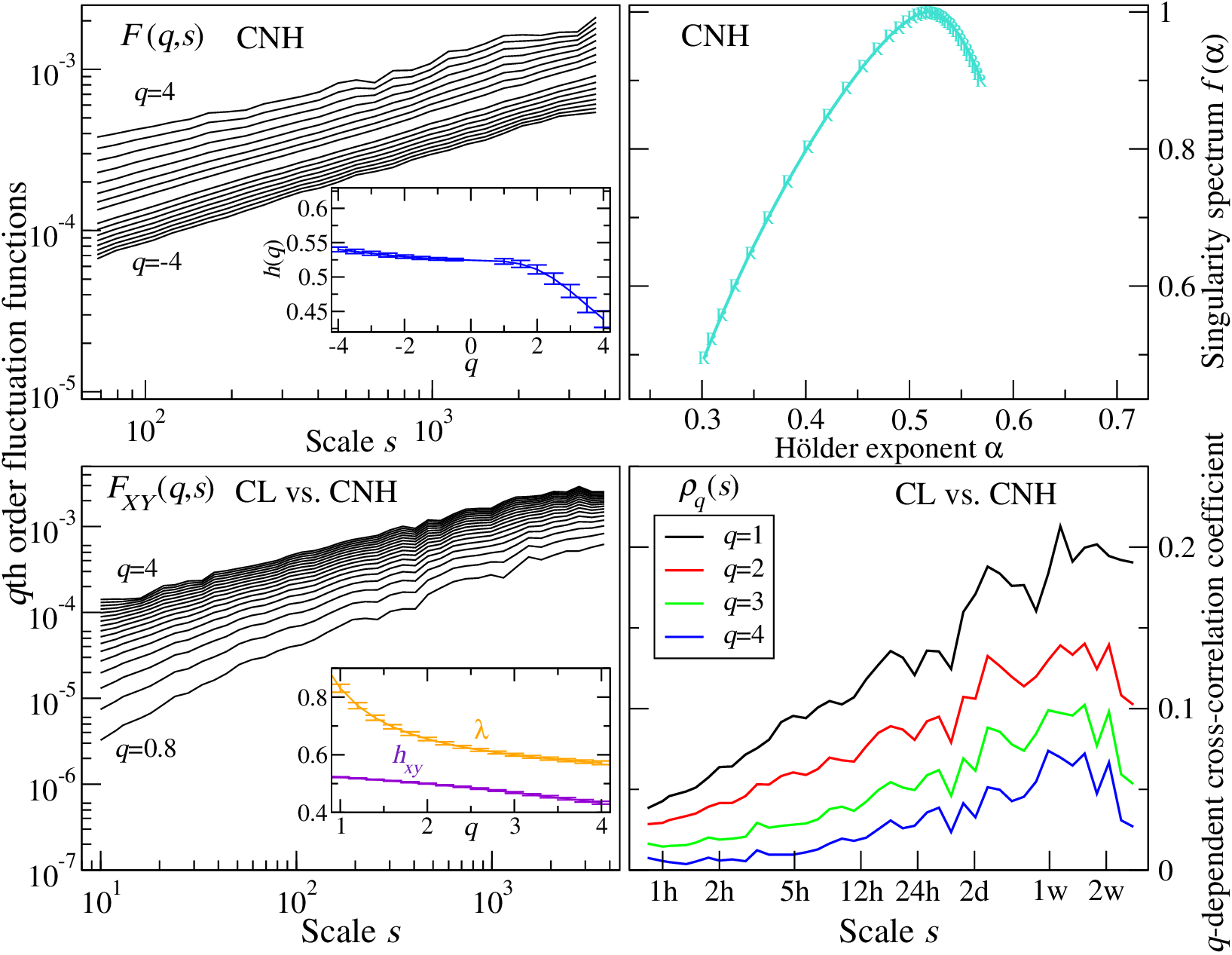}
\caption{Color online) The upper two panels display the family of $q$th-order fluctuation functions $F(q,s)$ for different values of $q\in [-4,4]$ with step 0.5 (the topmost one represents $q=4$) calculated for CNH/USD with the corresponding singularity spectra $f(\alpha)$. The lower-left panel displays the family of $q$th-order fluctuation cross-covariance functions $F_{XY}(q,s)$ for different values of $q$ with the step of 0.2 calculated for WTI Crude Oil futures (CL) versus CNH/USD. The topmost lines correspond to $q=4$ and the lowest one to $F_{XY}(q,s)$ that still develops a relatively regular shape with the corresponding limiting $q$-value explicitly listed. Inset illustrates the resulting $q$-dependence of $\lambda(q)$ versus the average of the generalized Hurst exponents $h_{xy}(q) = (h_x(q)+h_y(q))/2$ of the two series $x(i)$ and $y(i)$ under study. The lower-right panel displays the $q$-dependent detrended cross-correlation coefficient $\rho_q$ between CL and CNH/USD as a function of the temporal scale $s$ for $q=1$, $q=2$, $q=3$ and $q=4$.}
\label{fig:CNH}
\end{figure}
As it already has been revealed in Section \ref{data} the Chinese currency offshore CNH rate to USD followed rules of trading in several aspects distinct from the other currencies over the period considered here. The CNH multiscaling and its oil related characteristics are collected in Fig.~\ref{fig:CNH}. The CNH/USD fluctuation functions are seen to develop scaling and the resulting singularity spectrum has the width $\Delta_{\alpha}=0.27$, thus comparable to majority of the currencies of Fig.~\ref{fig:singspektrum}. It however is strongly left-sided asymmetric and the corresponding $A_{\alpha}=0.64$ is the largest among the instruments here considered. This indicates that multifractal organization of CNH/USD fluctuations significantly degrades for their smaller amplitudes. As a result, its cross-correlation to CL, as expressed by $F_{XY}(q,s)$, also develops scaling down in $q$ to about $q=0.8$, similarly as for EUR, GBP, JPY or PLN of Fig.~\ref{fig:FqCL}. Attention needs however to be drawn to the fact that absolute magnitudes of $F_{XY}(q,s)$ in the case of CNH on average are at least a factor of two smaller than those of the other currencies. Another manifestation of this fact, in view of Eq.~(\ref{eq:ca1}), is the difference $d_{xy}(q)$ between $\lambda(q)$ and $h_{xy}(q)=h_x(q)+h(q)/2$ shown in the inset to the lower-left panel of this figure. This difference is significantly larger (note different scales) than in any of the cases of Fig.~\ref{fig:FqCL} thus indicating weaker cross-correlations on average. Finally, these larger values of $d_{xy}(q)$ find their reflection also in a faster increase of $\rho_q(s)$ with increasing $s$ for CNH/USD versus oil as seen in the lower-right panel of Fig.~\ref{fig:CNH}. In spite of the faster increase this $\rho_q(s)$ still remains small as compared to the values of Fig.~\ref{fig:pDCCACL}. 

\subsection{Analysis in sub-periods}
\label{subperiods}
\begin{figure}[h!]
\centering 
\includegraphics[scale=0.45]{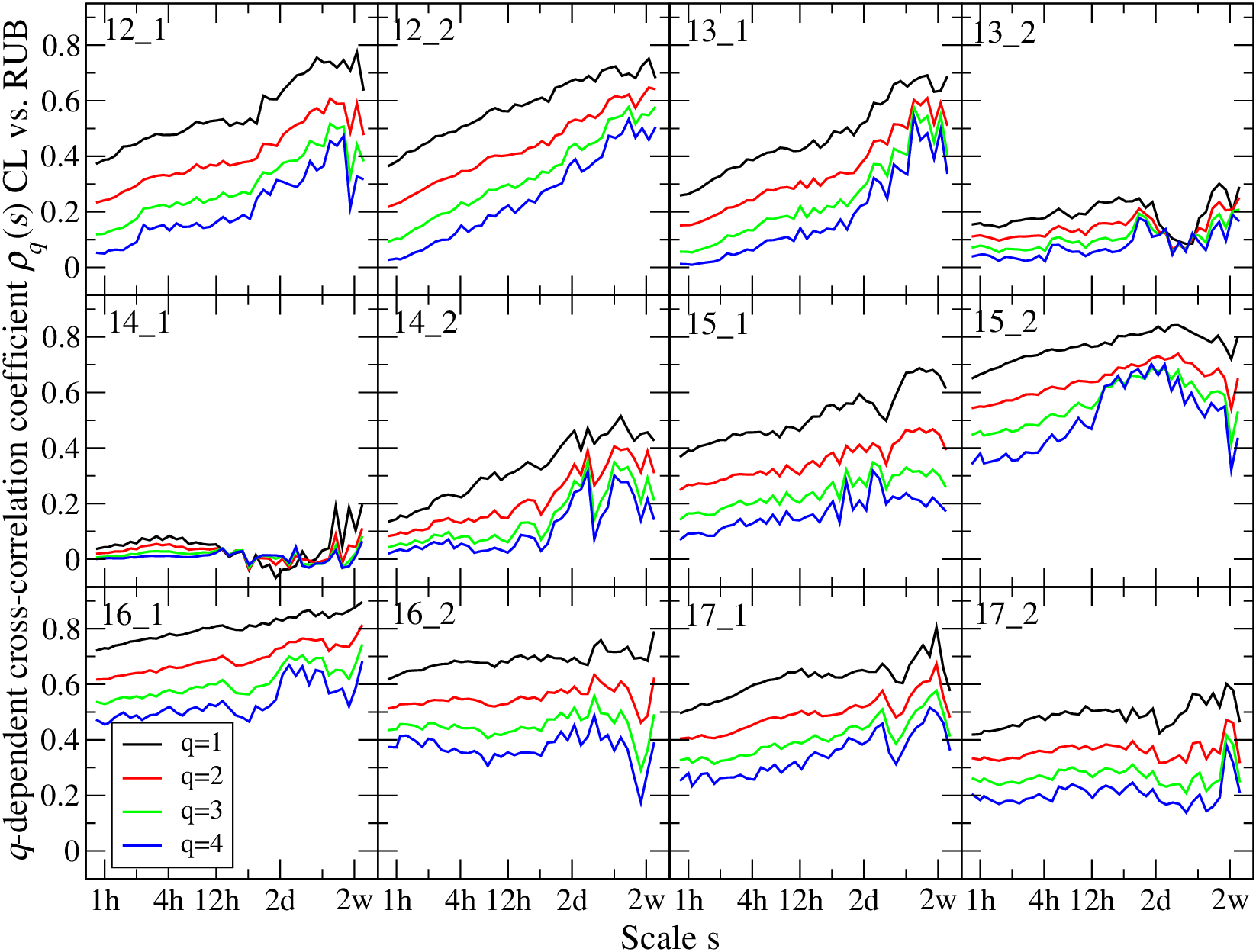}
\caption{(Color online) The $q$-dependent detrended cross-correlation coefficient as a function of the temporal scale $s$ $q=1$, $q=2$, $q=3$ and $q=4$, calculated between WTI Crude Oil futures (CL) and Russian ruble expressed in USD  in 12 sub-periods starting in 1st half of 2012 until 2nd half of 2017.} 
\label{fig:pq_CLRUBsub}
\end{figure}
As Fig.~\ref{fig:CL} shows the 6 years period here considered displays variety of trends both on the oil market as well as on the currency markets. Such changes are known \citep{Drozdz2018} to influence the multiscaling characteristics of correlations among assets therefore in the following a more local analysis of cross-correlations is performed by splitting the entire period into the disjoint half-year sub-periods. Since the Russian ruble belongs to those currencies that develop the strongest correlations to the oil prices it is this currency whose example is in detail displayed in Fig.~\ref{fig:pq_CLRUBsub} for the same four $q$-values as in Fig.~\ref{fig:pDCCACL}. As it is clearly seen the magnitude of correlations appears to depend on time. This dependence is particularly strong in the vicinity of the sizeable decline on the oil market that started in mid 2014 and the oil prices dropped down by more than a factor of two by the end of that year. It is remarkable that just before this decline the corresponding ruble correlation coefficients $\rho_q(s)$ reached almost a level of null signalling an entire decorrelation of the two assets. During the oil price decline its correlation to the ruble started systematically resuming and even exceeded the level of early 2012 at the turn of 2015/2016 when the oil price reached the minimum value during the period here considered. It is also interesting to see that in all these 12 sub-periods the ordering of the $\rho_q(s)$-values in $q$ is always preserved, i.e., they are the largest for $q=1$ and systematically decrease when going to $q=4$. 

\begin{figure}[h!]
\centering 
\includegraphics[scale=0.45]{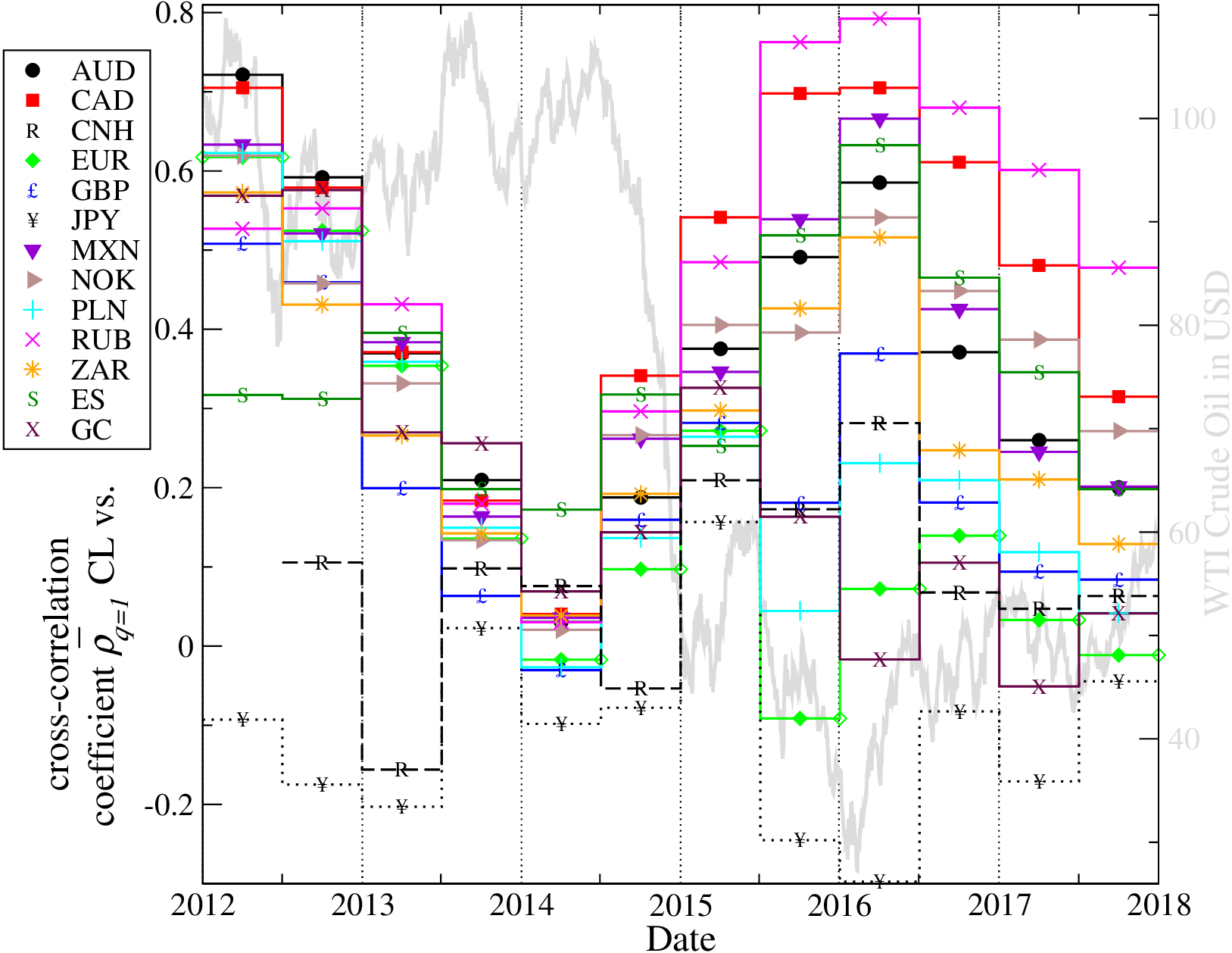}
\caption{(Color online) The $q=1$ detrended cross-correlation coefficient averaged on time scales $\overline{\rho}_{q=1}$ in 12 sub-periods since 1st half of 2012 until 2nd half of 2017 between WTI Crude Oil futures (CL) and other financial instruments.}
\label{fig:oilall}
\end{figure}
Similar dependences as for the ruble appear to apply for AUD, CAD, NOK and to some extent also for MXN. The GBP and especially EUR reveal much weaker correlation to the oil trend changes. As already noticed above, the JPY develops an independent oil related dynamics and its correlation is often negative. CNH remains weakly correlated to oil and in the first half of 2013 it even shows negative correlation thus resembling the JPY. Later on however it assumes behaviour similar to EUR and GBP. Interestingly, during the first three (2012--2014) years of the time period here considered the gold correlation to oil follows the behaviour of oil related currencies while in the later three (2015--2017) years it only shows a small fraction of such an effect. The world largest stock market index, the S\&P500, on the other hand, remains sizeably oil correlated over the whole time-span considered though interestingly, this correlation also reaches minimum just before the oil price strong decline in mid 2014. All these facts are globally illustrated in Fig.~\ref{fig:oilall} which shows the averages $\overline{\rho}_{q=1}$ of $\rho_{q=1}(s)$ over the corresponding range of time scales $s$.

\section{Asymmetry in cross-information flow}
\label{causation}

Multiscale cross-correlations are caused by subtle effects and they quickly disappear when the series are desynchronized as shown in \cite{oswiecimka2014}. The formalism of Eqs. [\ref{Fxy} - \ref{rhoq}] allows to get some insight into possible asymmetry effects in the information flow that determines the character of cross-correlations between the two time series and, in particular, some advances or retardations. This can be studied by shifting the series relative to each other and repeating the procedure of quantifying the cross-correlations. An example of the scale $s$ dependence of $\rho_q$ $(q=1,2,3,4)$ coefficient for the Russian ruble versus oil for the three variants of relative correspondence  between their times series is shown in Fig.~\ref{fig:CLRUBcauseq}. The solid lines represent $\rho_q(s)$ for their original, synchronous alignment. The dashed lines are obtained by shifting the oil price series one step forwards relative to the one of ruble. This thus quantifies the degree of correlations between the oil price change at a particular instant of time with the one of ruble 5 min later. The dotted lines, on the other hand, reflect the result of opposite shift, i.e., the ruble price series is shifted by one step (thus 5 min) forwards relative to the one of oil. A strong weakening of correlations due to such shifts is seen at the smaller, hourly scales ($s$) of detrending. One particularly interesting effect to be noticed, however, is that the former (dashed line) case systematically preserves more cross-correlation than the latter (dotted line). This seems to indicate that these are the oil price changes that more influence the ruble and dictate the information flow from the oil to ruble than vice versa. On the larger, weekly scales of $s$ such shifts are becoming gradually ineffective. 

\begin{figure}[h!]
\centering 
\includegraphics[scale=0.45]{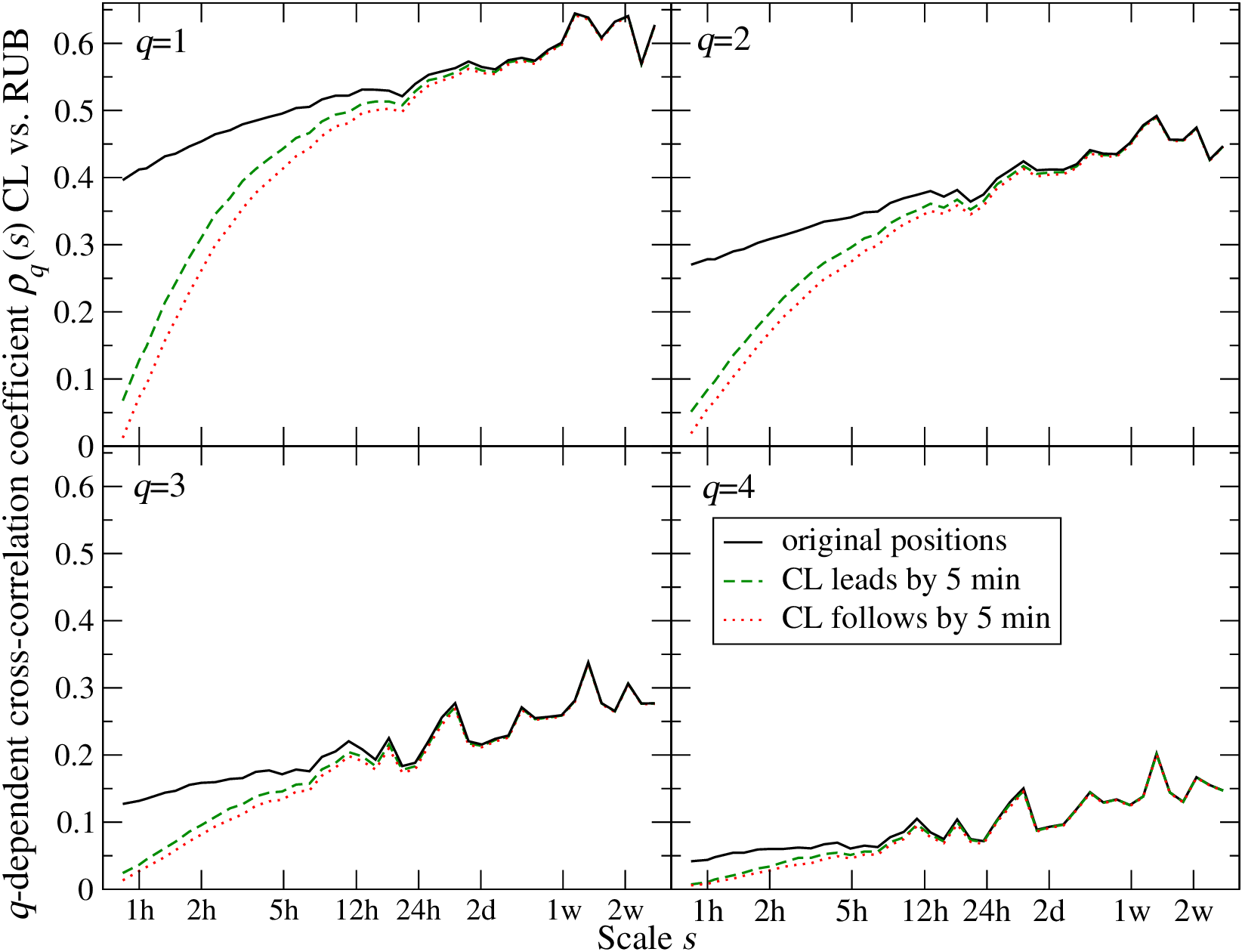}
\caption{(Color online) The $q$-dependent detrended cross-correlation coefficient calculated for different synchronization levels of the time series representing WTI Crude Oil futures (CL) and Russian ruble expressed in USD (RUB): (1) the synchronous (original) index positions, (2) RUB retarded by 5 min with respect to CL (CL leads by 5 min), (3) CL retarded by 5 min with respect to RUB (CL follows by 5 min) ($q=1,2,3,4$).}
\label{fig:CLRUBcauseq}
\end{figure}
As Fig.~\ref{fig:CLRUBcauseq} also shows it is $\rho_{q=1}(s)$ for which the above effects are seen to be the strongest. Therefore, the result of analysis in the half-year sub-periods, analogous to Fig.~\ref{fig:pq_CLRUBsub}, of this particular $\rho_{q=1}(s)$ is displayed in Fig.~\ref{fig:oil-rubel-subperiods}. The dominant effect of driving the ruble price by the one of oil appears to occur in the 2nd half of 2014, when the largest fall in the oil price took place. One also needs to keep in mind in this context that the frequency of price changes available in the present study is 5 min which in the contemporary markets - from the perspective of the speed of information processing - is already a relatively long period of time. It is thus very likely that the observed effects of the asymmetry may appear in an even more pronounced form on the shorter time-scale, say of the order of seconds. 

\begin{figure}[h!]
\centering 
\includegraphics[scale=0.45]{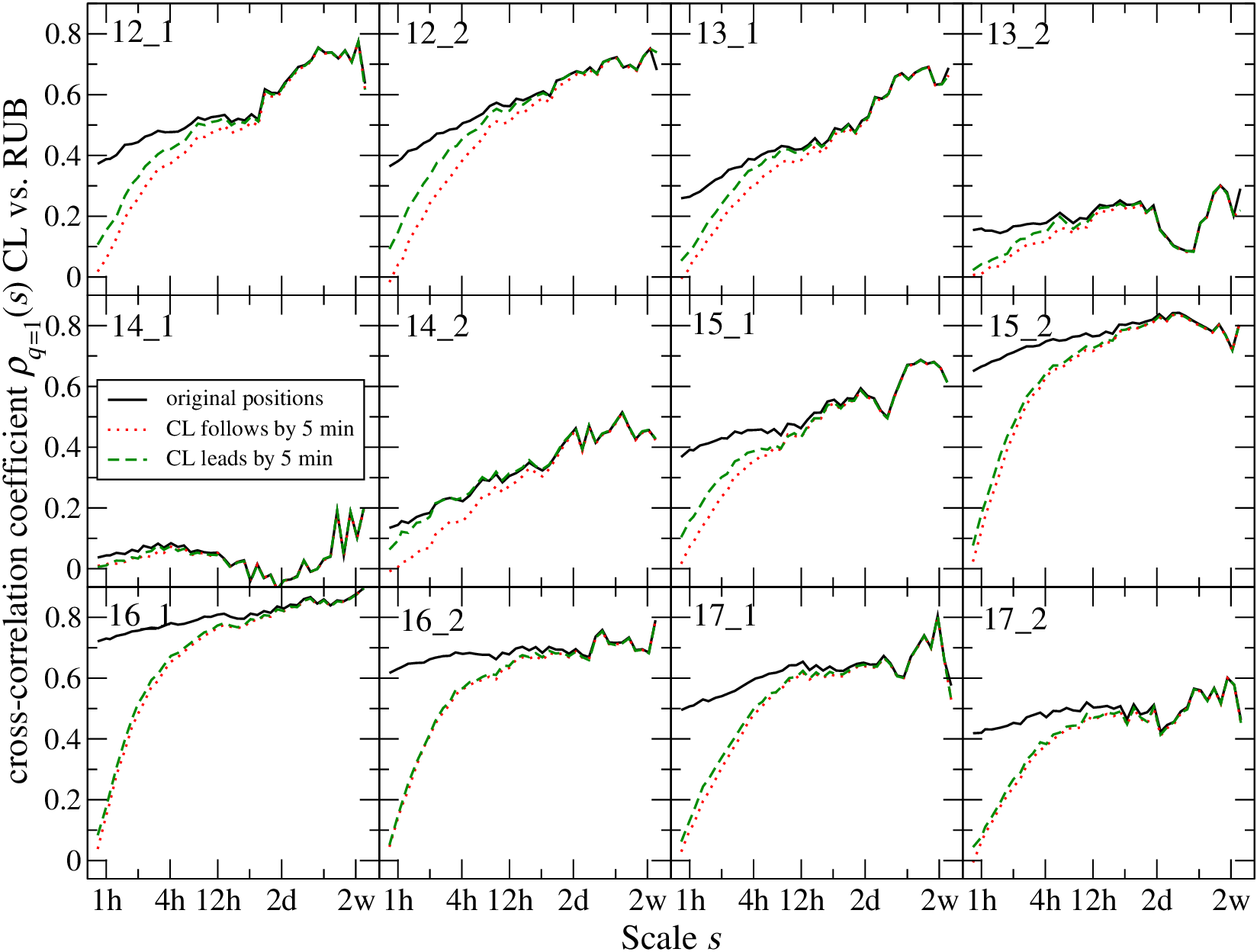}
\caption{(Color online) The $q=1$ detrended cross-correlation coefficient calculated in 12 sub-periods since 1st half of 2012 until 2nd half of 2017 for three different synchronization levels of the time series representing WTI Crude Oil futures (CL) and Russian ruble expressed in USD (RUB): (1) the synchronous (original) positions, (2) RUB retarded by 5 min with respect to CL (CL leads by 5 min), (3) CL retarded by 5 min with respect to RUB (CL follows by 5 min).}
\label{fig:oil-rubel-subperiods}
\end{figure}

Finally, Fig.~\ref{fig:oil-all} collects the results of $\rho_{q=1}(s)$ for cross-correlations - as before direct, advanced and retarded - between oil and all other financial instruments here considered. 
In addition to the Russian ruble already discussed the effect of similar asymmetry versus oil can be seen for the South African rand (ZAR) and some trace of it for Mexican peso (MXN) and Norwegian krone (NOK), and, interestingly, in the opposite direction (oil retarded) in Euro (EUR) and in the gold (GC). 

\begin{figure}[h!]
\centering 
\includegraphics[scale=0.45]{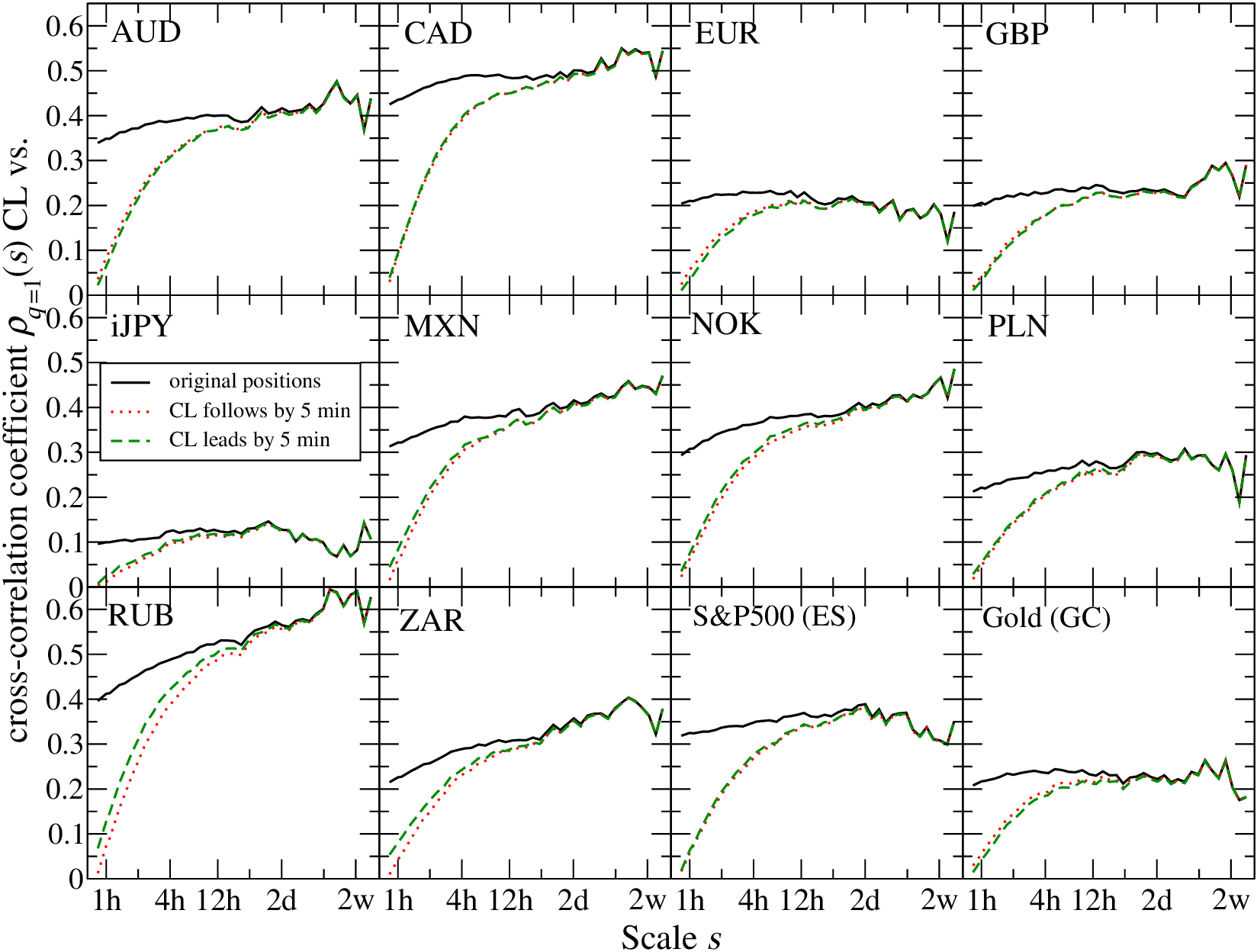}
\caption{(Color online) The $q=1$ detrended cross-correlation coefficient between WTI Crude Oil futures and the 10 currencies, S\&P500 futures and gold futures since 1st half of 2012 until 2nd half of 2017 for three different synchronization levels of the time series representing WTI Crude Oil futures (CL) vs. rest instruments in USD: (1) the synchronous (original) positions, (2) RUB retarded by 5 min with respect to CL (CL leads by 5 min), (3) CL retarded by 5 min with respect to RUB (CL follows by 5 min).}
\label{fig:oil-all}
\end{figure}

\section{Summary}
\label{Concl}
The present study demonstrates applicability of the formalism of multifractal detrended cross-correlation analysis (MFCCA) to investigate cross-correlations between WTI Crude Oil futures versus currencies expressed in USD, gold futures and S\&P500 futures. The related methodology proves to provide the appropriate formal tools to detect various subtle effects both in the temporal organization of fluctuations of the individual time series representing the price changes as well as cross-correlations among them including the nonlinear ones resulting in multifractality. From the perspective of multiscaling the analysed time series appear to be multifractally organized developing the left-sided asymmetry of the singularity spectrum, however, thus indicating that this multifractality is mainly due to medium and larger size fluctuations. The S\&P500 futures constitutes an exception in this respect as its singularity spectrum is almost symmetric. As the cross-correlation analysis between pairs of different financial instruments shows this multifractal organization often tends to synchronize among them. The related detrended cross-correlation coefficient $\rho_q(s)$ analysis shows that the strongest cross-correlations of this kind take place between oil on the one side and either the S\&P500 futures or the currencies of oil producing countries on the other. 
The half-year sub-period analysis reveals that in the entire time period studied 2012-2017 the degree of cross-correlations increases systematically during the bear phase on the oil market and it saturates after the trend reversal in 1st half of 2016. Very interestingly, the beginning of oil price decline in mid 2014 is seen to be proceeded by an almost entire decorrelation from oil of all the instruments considered. This effect as a potential forecasting signal definitely deserves a more systematic verification.
Finally, the MFCCA methodology appears applicable towards getting some signatures of possible causal relations between the considered observables. Analysis of the asymmetry in information flow in cross-correlation characteristics estimated for different variants of synchronization of the time series indicates dependence of the Russian ruble rate on the oil prices. The strongest causality effect is observed in the 2nd half of 2014, when the largest fall in the oil prices took place. From a practical perspective such effects can potentially be exploited by the market participants in shaping the trading strategies. They may also be of great value for improving and extending the financial forecasting models explicitly incorporating multifractality like the Multifractal Model of Asset Returns \citep{mandelbrot1997,calvet2002,oswiecimka2006a,lux2008}.

\newpage
\bibliography{CLartfinalporec}

\end{document}